\RequirePackage[columnwise]{lineno} 
\documentclass[journal=jacsat,manuscript=article]{achemso}

\usepackage{amsmath}           % Equation module - AMS math package
\usepackage{graphicx}          % Figure module - graphicx package
\usepackage[version=3]{mhchem}
\usepackage{dcolumn}           % Align table columns on decimal point
\usepackage{xcolor}
\usepackage{ulem}

\author{Jae Whan Park}		    % Author name
\affiliation{Center for Artificial Low Dimensional Electronic Systems, Institute for Basic Science (IBS), Pohang 37673, Republic of Korea}
\author{Hyeonjung Kim}
\affiliation{Center for Artificial Low Dimensional Electronic Systems, Institute for Basic Science (IBS), Pohang 37673, Republic of Korea}
\altaffiliation{Department of Physics, Pohang University of Science and Technology (POSTECH), Pohang 37673, Republic of Korea}
\author{Han Woong Yeom}
\email{yeom@postech.ac.kr}	    % E-mail address for corresponding author
\affiliation{Center for Artificial Low Dimensional Electronic Systems, Institute for Basic Science (IBS), Pohang 37673, Republic of Korea}
\altaffiliation{Department of Physics, Pohang University of Science and Technology (POSTECH), Pohang 37673, Republic of Korea}

\title{Disentangle Intertwined Interactions in Correlated Charge Density Wave with Magnetic Impurities}      

\begin{document}               % Initiate the document

\begin{abstract}				% Abstract
Magnetic impurities in strongly correlated electronic systems serve as sensitive probes to a wide range of many-body quantum phenomena. Broken symmetries in such a system can lead to inequivalent lattice sites, and magnetic impurities may interact selectively with particular orbitals or sublattices. 
However, the microscopic mechanisms behind such site-specific interactions have been poorly understood.
Here, we explore the behavior of individual Fe adatoms on a cluster-Mott charge-density-wave (CDW) system of 1T-TaS$_2$ utilizing scanning tunneling microscopy/spectroscopy (STM/STS) and density functional theory (DFT). Our measurements uncover pronounced site-dependent electronic states of CDW clusters with Fe adatoms, indicating distinct local coupling to cluster-Mott states. DFT calculations identify three distinct types of interactions; hybridization with localized correlated electrons, distorting the CDW cluster, and charge transfer. In particular, the hybridization of Fe $3d$ and half-filled Ta $5d_{z^2}$ orbitals suppresses the Mott insulating state for an adatom at the center of a CDW cluster. While the results underscore a crucial role of the direct orbital hybridization and the limitation of the prevailing single-site Kondo impurity model, they suggest the possibility of controlling entangled interactions separately in a cluster Mott insulator. 
\end{abstract}

\noindent{\fontfamily{phv}\selectfont \textbf{Keywords}}
STM/S, DFT, magnetic impurity, charge density wave, correlated electron, 1T-TaS$_2$

\

%%%%%%%%%%%%%%%%%%%%%%%%%% Introduction %%%%%%%%%%%%%%%%%%%%%%%%%%
% Nonequilibrium electronic phase

\noindent{\fontfamily{phv}\selectfont \textbf{INTRODUCTION}}

%Interest in Magnetic impurity on correlated systems
The coupling of spin and charge degrees of freedom in strongly correlated electron systems lies at the core of current condensed matter physics research as manifested in quantum spin liquids, heavy fermion behaviors, and unconventional superconductivity \cite{lawa17, klan17, chen22, shen22, sipo08, ang12,vano21,crip24,bala06}. In particular, magnetic impurities can interact in exotic ways not only with host spins but also with charge and orbital degrees of freedom of correlated host electrons. Such interactions can profoundly reshape local and global electronic structure as observed in, for example, Yu–Shiba–Rusinov states in superconductors, Kondo screening in heavy fermion systems and Mott insulators, and impurity-induced reconstruction of charge-density-wave (CDW) or orbital-ordered states in transition metal dichalcogenides and cuprates \cite{yazd97,nadj14,khal95,vojt06,spin15,faze80,hana09,kohs12}.
A prototypical setting to explore such effects has been magnetic impurities in Mott insulators, where the subtle balance between localized spins and charge gaps provides a fertile ground for unconventional impurity physics, including interactions with collective spin excitations such as magnons \cite{batt17,miya18,sing98}.

The interaction of magnetic impurities with a Mott insulator gains a new facet in cluster-based Mott insulators, which have multiple sites within each unit cell. 
In this type of Mott insulators, identified in GaTa$_4$Se$_8$, Nb$_3$Cl$_8$, Mo$_3$O$_8$, and Kagome materials \cite{magn24,niko21,hu23,chen18,zhen24,yang25}, the combination of strong electron correlation with complex local structures can produce rich correlation phenomena including flat band physics and spin frustration. 
Not limited to bulk crystals, the cluster-Mott-insulator idea is crucial in understanding correlated insulating states in Moir{\' e} superstructures of graphene \cite{cao18,chen2019,chen21} and transition metal dichalcogenides \cite{fei22,naka16,Liu21}. 
The multi-site effect due to the cluster nature can readily be manifested in the interaction with magnetic impurities. 
For example, an exotic interaction of a magnetic impurity with gap-less spinons, such as the spinon Kondo effect \cite{chen22}, was introduced recently in the triangular Mott-CDW insulator of 1T-TaSe$_2$ (or 1T-TaS$_2$ and 1T-NbSe$_2$) \cite{ruan21,zhang24,zheng24,he22}. 
At the same time, strongly site-dependent electronic states of magnetic impurities were noticed \cite{chen22}, whose origin are elusive. 
The existence of multiple sites goes beyond the theoretical models utilized, which adopt single-site Hamiltonians without any details of local impurities and their site-specific interactions \cite{ding24,pang25}, to prevent understanding of local impurity-host interactions. 

%In this paper...
In this study, we investigate magnetic impurities Fe adsorbed on the correlated CDW material 1T-TaS$_2$ using a combination of scanning tunneling microscopy/spectroscopy (STM/STS) and density functional theory (DFT) calculations. 
STM/STS measurements reveal three different Fe adsorption sites, which exhibit distinct electronic behaviors, the suppression of the Mott-Hubbard state, the electron doping, and the emergence of in-gap states, respectively. 
Our DFT calculations closely reproduce these experimental findings and provide microscopic insight into the site-dependent interactions underlying the electronic modifications. We identified three key mechanisms depending on sites: (i) the direct $d$ orbital hybridization involving Mott-Hubbard states on the cluster center, (ii) a weak charge-transfer interaction on the cluster edge, (iii) the interaction with the CDW structure on the site between clusters. 
These findings highlight the critical role of local chemical interactions in determining the electronic structure and call for a reassessment of the previous interpretations based solely on many-body effects such as spinon excitations and Kondo phenomena \cite{chen22}. On the other hand, the present result suggests the possibility for manipulating quantum properties of cluster Mott insulators through defect engineering. 

\vspace{1em}

\noindent{\fontfamily{phv}\selectfont \textbf{RESULTS}}

1T-TaS$_2$ (also 1T-TaSe$_2$, 1T-NbS$_2$, or 1T-NbSe$_2$) is a prototypical quasi two-dimensional (2D) crystal, which consists of hexagonal Ta (or Nb) layers octahedrally coordinated by S (or Se) atoms.
Below 180 K, it undergoes a transition into a commensurate CDW phase, forming characteristic David-star (DS) clusters that constitute a $\sqrt{13} \times \sqrt{13}$ CDW superstructure [Fig. 1(h)] \cite{brou80}.
Each CDW cluster has thirteen Ta atoms and twelve of them shrink toward the central Ta atom for the CDW formation. 
Among the thirteen Ta $5d$ electrons in each CDW cluster, one unpaired $d_{z^2}$ electron becomes localized at the central Ta atom to form a Mott insulating state in the monolayer limit \cite{faze79}, while the remaining twelve electrons contribute to the gapped CDW bands [Fig. 1(i)].
In the bulk, the stacking order favors an alternating bilayer configuration known as AC stacking \cite{rits18,lee19}. 
Each bilayer exhibits a band-insulating character due to interlayer bonding of the unpaired $d_{z^2}$ electrons at cluster centers [Fig. 1(j) and Supplementary Fig. 1] \cite{rits18,lee19, butl20}.
This bonding-antibonding-type band gap (about 0.4 eV) was reported to prevail on the surfaces observed \cite{butl20, wu22}, while there can be various different types of band gaps depending on the stacking order and the interlayer spin ordering near the surface region \cite{park23}.  
The bonding-antibonding gap in the bilayer structure is, however, not totally trivial, since the system still has a substantial local magnetic moment, as shown in Fig. 1(j), which is more enhanced in the surface bilayer than in the bulk \cite{park23}. 
Nevertheless, as revealed in the following DFT calculations, the adsorption behaviors of Fe adatoms on a bilayer and a monolayer are consistent since the adsorption on the top layer effectively breaks the interlayer coupling, through electronic effects, structural effects, or a combination of both, depending on the adsorption sites (Supplementary Note I).  
Based on this, we use the terminology of Mott-gap and low(upper)-Hubbard bands L(U)HB for simplicity in the following discussion.

\vspace{1em}
%{\bf Three distinct Fe adsorption configurations.}
Figure 1(a) shows typical STM topographic images of the commensurate CDW phase of 1T-TaS$_2$ at 4.5 K with a small number of Fe adsorbates.
The array of medium-contrast protrusions in a triangular shape correspond to DS CDW clusters, while brighter ones represent Fe adatoms [circles in Fig. 1(a)].
We identified mainly three distinct adsorption geometries, which are shown in Figs. 1(b), 1(c), and 1(d), respectively. 
A Fe adatom can sit symmetrically at the center of a DS cluster, at one tip of a DS cluster, and in between two neighboring DS clusters. 
These sites are called on-center, on-edge, and off-cluster sites, respectively, and consistent adsorption sites were observed in the previous STM studies for Co and Fe adatoms on 1T-TaSe$_2$ and 1T-TaS$_2$, respectively \cite{chen22, fuji03}.
These three sites encompass eight symmetry-inequivalent hollow sites in the surface S layer; three with Ta atoms underneath (A, B, and C in Supplementary Fig. 3 and 4) and the other five without (H1-H5 in Supplementary Figs. 3 and 4).

\vspace{1em}
%{\bf Site-dependent electronic structures.}
Although there are marginal variations, the change of electronic states upon Fe adsorption can be reasonably categorized into three types of sites mentioned above [Fig. 1(e)-1(g) and Supplementary Fig. 2 and Fig. 5].
For the on-center adsorption, both LHB and UHB states disappear, while a pronounced peak emerges between $-0.5$ to $-0.3$~eV range as shown in the STS map of Fig. 1(e) (open triangle) and the representative STS spectra in Fig. 2(a). 
In clear contrast, an on-edge and an off-cluster adatoms produce in-gap states below the upper-Hubbard band (UHB) [Fig. 1(f)] and just above the Fermi level [Fig. 1(g)], respectively. 
One can notice that the in-gap state of the on-edge adatom is formed by a rigid downward shift of LHB and UHB [see green arrows in Fig. 1(f) and Fig. 3(a)]. 
However, the in-gap state on the off-cluster site is not consistent with the rigid energy shift with the band gap greatly reduced [see Fig. 1(g) and Fig. 4(a)]. 
Note also an upward band bending in the neighboring unit cells in this case, which is opposite to the case of the on-edge adsorption. 

\vspace{1em}
%{\bf DFT model and energetics.}
The origins of the site-dependent spectral features observed are understood by DFT calculations based on a 1T-TaS$_2$ bilayer. 
Our calculated LDOS for a bare bilayer structure reproduces the experimental STS features well [Fig. 1(j) and Supplementary Fig. 5], aside from the underestimation of the band gap by about 0.1 eV.
This is attributed to the well-known limitation of the conventional DFT approach in reproducing band gaps in fully gapped systems~\cite{perd83,sham83}.
We used a $\sqrt{3} \times \sqrt{3}$ CDW supercell based on a bilayer structure [Fig. 1(j)], which incorporates a single Fe adatom among three CDW unit cells on the top layer to avoid possible adatom-adatom interactions.
To capture strong electron correlation effects for a local Fe adatom, we employed the DFT+$U$ method [see Fig. 1(j)] with the on-site Coulomb energy $U$ of 2.3 eV added.
Otherwise, DFT calculations predict metallic ground states for Fe-adsorbed unit cells, deviating qualitatively from the experimental observations (Supplementary Fig. 5).

Our DFT+$U$ calculations with Fe adatoms yield adsorption energies of $-3.455$~eV, $-3.931$~eV, and $-3.892$~eV for the on-center, on-edge, and off-cluster sites, respectively (see also Supplementary Fig. 3). 
These energetics are in qualitative agreement with the experimentally observed site populations; 63, 86, and 84 counts, respectively for on-center, on-edge, and off-cluster sites for a surface area of 180 $\times$ 180 nm$^2$.
The large adsorption energies indicate that Fe atoms participate in strong chemical interactions with the surface.
This energy scale exceeds substantially those of many-body interactions such as spin excitations or Kondo couplings, indicating unambiguously that direct chemical interactions would dominate the local electronic properties over possible many-body effects.

\vspace{1em}
%{\bf Strong hybridization between a magnetic adatom and a correlated electron.}
More details of the adatom-substrate interactions are revealed by analyzing LDOS.
For the on-center site, calculated LDOS [Fig. 2(b)] shows the full suppression of the LHB/UHB states.
This suppression is fully consistent with the STS observation [Figs. 1(e) and 2(a)], which is also visualized in the STS spatial map in Fig. 2(e) for the LHB state (see also Supplementary Fig. 6).
Instead, a localized state emerges at the Fe adatom site (red line) below the energy of the pristine LHB band in the calculation (-0.23 eV). 
This state is consistently observed in the experiment but at a slightly lower energy of -0.34 eV below the Fermi energy [open triangle in Fig. 2(a)].
This LDOS weight comes partly from the central Ta $5d_{z^2}$ orbital, which is responsible for the LHB state of the pristine unit cell [the green curve in Fig. 2(b)], and partly from Fe 3$d$ orbitals [red curve in Fig. 2(b)].
This result clearly indicates the strong hybridization between these two $d$ orbitals (see Supplementary Figs. 7 and 8).
This state is the bonding part of the d-d hybridization, and the antibonding state is located at about 1 eV higher in energy (see Supplementary Fig. 7) due to the large exchange splitting of Fe $3d$ levels. 
The magnetic moments of the Fe atom and the central Ta atom are calculated to be 2.59 and 0.16 $\mu_B$, respectively, with their spins aligned in opposite directions. 

Despite the strong local electronic modification, the CDW structure remains almost intact, as shown in the fully relaxed atomic structure [bottom panel of Fig. 2(d)]. 
This is because the on-center Fe adatom interacts selectively with the Ta $5d_{z^2}$ correlated electron at the center of a CDW cluster, which is not involved in the CDW formation. 
We note further that the CDW unit cell underneath the Fe-adsorbed top layer unit cell restores the single-layer Mott insulating behavior as the interlayer coupling is substantially weakened by the Fe adsorption on the top layer (Supplementary Fig. 9). 

\vspace{1em}
%{\bf Charge transfer for a weakly interacting adatom.}
The on-edge adsorption configuration [Fig. 1(c) and Fig. 3] exhibits markedly different STS spectral features from those of the on-center case.
At the center of a CDW cluster with a Fe adatom on its edge, the STS spectrum exhibits a rather rigid shift to a lower energy with a marginal reduction of the Mott gap [green line in Fig. 3(a)] and the partial reduction of the UHB peak.
The reduction of the Mott gap and the UHB intensity is well expected from the electron doping into the Mott insulator, which is indicated by the shift of the LHB/UHB state \cite{jian19}. 
On the Fe adatom site, a similar rigid shift is observed but with the strong lowest unoccupied state at 0.1 eV. 
The theoretical LDOS in Fig. 3(b) closely reproduces both the downward shifts of UHB/LHB (green line) on the cluster center and the prominent unoccupied-state peak associated with the Fe atom.
The spatially resolved LDOS from the $dI/dV$ map (Supplementary Fig. 6) indicates that the Fe 3$d$ state is responsible for the LDOS peak at 0.1 eV. 

Structurally, the Fe adsorption leads to a small (approximately 0.04 $\sim$ 0.09~\AA) elongation of the nearest-neighbor Ta–Ta bond lengths within the CDW cluster.
However, the resulting bond lengths (3.223 $\sim$ 3.371~\AA) remain shorter than the ideal 1$\times$1 lattice constant of 3.374~\AA.
That is, the DS cluster experiences only minor structural perturbations to preserve its overall CDW distortions [Figs.~3(b) and 3(d)]. 
The structural perturbation on the CDW cluster with Fe, however, reduces the CDW band gap by 0.05 eV in the DFT+U calculation [Fig.~3(c)]. %0.254and 0.3%
The reduction of the CDW gap, in turn, makes the LHB overlap partly with the valence band top, which is believed to induce the charge transfer into the Mott bands. 
The LDOS change is visualized by the Fe-induced charge redistribution map [bottom panel in Fig. 3(d)].
In summary, the on-edge adatom modifies marginally the CDW cluster, which induces small charge transfer into the Mott bands without destroying the Mott state themselves.

\vspace{1em}
%{\bf Coupling with charge density wave order.}
In the case of the off-cluster adsorption, distinct from the two above configurations, the Fe adatom gives rise to a pronounced in-gap state within the Mott gap [Fig. 4a] and the substantial perturbation of the CDW unit cell. 
When the Fe atom adsorbs at a hollow site between adjacent DS clusters, it exerts an attractive force on nearby Ta atoms, inducing a partial collapse of the CDW structure in the two neighboring clusters [inset in Fig. 4(b),  Fig. 4(d), and Supplementary Fig. 10].
In particular, one edge Ta atom is detached from its original DS cluster with its Ta–Ta bond length increasing to 3.507~\AA, and forms a bond with the neighboring cluster at a distance of 3.309~\AA~[Fig. 4(d)]. 
As a result, the original DS clusters convert to two distinct clusters composed of 12 and 14 Ta atoms, respectively. 
The corresponding charge redistribution extends beyond the CDW cell [bottom panel in Fig. 4(d)], indicating clearly that the local CDW order is broken.
Consequently, the CDW band gap is substantially reduced to 0.17 eV from its pristine value of 0.3 eV [see gray filled curve in Fig. 4(b) and 4(c)].
The in-gap state observed in this configuration is well reproduced in the DFT calculations. 
This state originates from the interplay of the reduced CDW gap mentioned above and the hybridization of the UHB state with Fe $3d$ orbitals.
As shown in Fig. 4(d), the UHB broadens substantially in energy [Fig.~4(c)] due to the hybridization with Fe $3d$, while the LHB retains its localized Ta-$5d_{z^2}$ orbital character.

% ---------- FIG 1: STM
\begin{figure}[t]
\centering{ \includegraphics[width=16 cm]{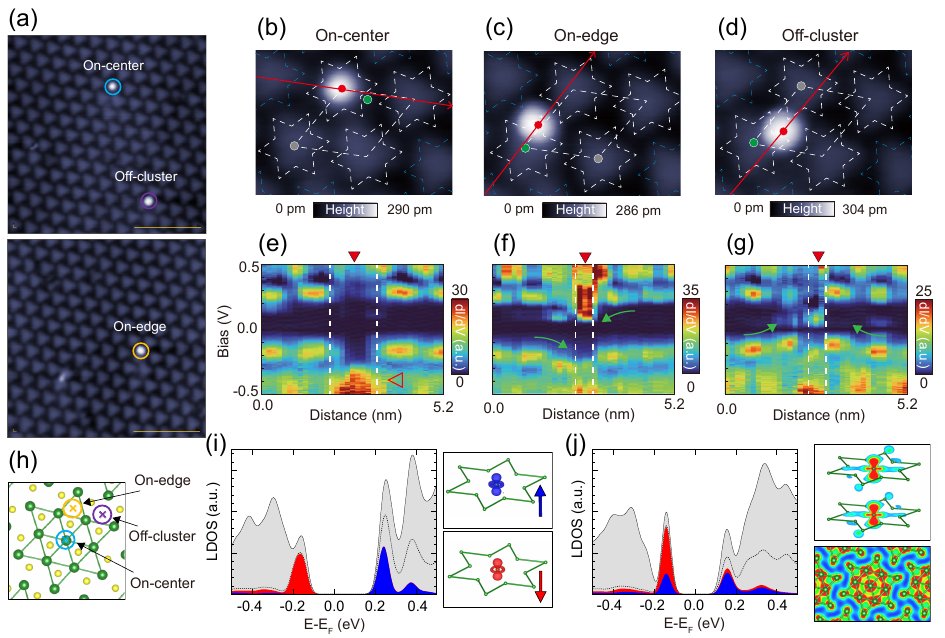} }
\caption{ \label{fig1}
STM images and STS line profiles for three distinct Fe adsorption configurations on 1T-TaS$_2$.
(a) Representative STM topography of the commensurate CDW phase in 1T-TaS$_2$ with Fe adsorbates ($V_b$ = -0.5 V, $I$ = 50 pA, scale bar = 5 nm). The periodic protrusions correspond to David-star (DS) clusters, while brighter features indicate adsorbed Fe atoms.
(b-d) Zoomed-in STM images showing Fe atoms at three characteristic adsorption configurations: (b) on-center, (c) on-edge, and (d) off-cluster sites.
(e-g) Corresponding STS line profiles acquired along the arrows in (b-d), across the Fe sites (see Supplementary Fig. 2).  
(h) Schematic illustration of a single DS cluster forming the $\sqrt{13} \times \sqrt{13}$ CDW superstructure. 
(i) Theoretical LDOS of pristine single-layer 1T-TaS$_2$. Gray-filled curve shows the total DOS from all thirteen Ta atoms within a DS cluster (scaled by 1/3 for clarity).
Dashed line represents the states localized at the central Ta atom, while red and blue filled curves represent spin-resolved contributions from the central Ta $5d_{z^2}$ orbital. 
Right: Spatial isosurfaces of $d_{z^2}$-like charge density corresponding to the lower and upper Hubbard bands, highlighting the Mott-localized states. The spin density is provided by color density.
(j) Bilayer 1T-TaS$_2$. Right: Charge density maps at the central Ta site, integrated over energy windows of $-0.2$ to $0.0$~eV (top) and $-1.0$ to $0.0$~eV (bottom, top view), illustrating $d_{z^2}$ orbital character and CDW-induced modulation in the bilayer system, respectively.
}
\end{figure} 

% ---------- FIG 2
\begin{figure}[t]
\centering{ \includegraphics[width= 8cm]{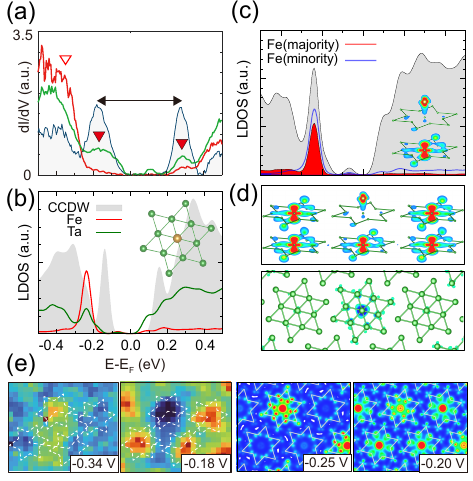} }
\caption{ \label{fig2}
On-center Fe adsorption coupled to surface-correlated electrons.
(a) Point STS spectra measured at the Fe adsorption site (red), a nearby pristine region (green), and a neighboring CDW center (gray filled),corresponding to the respectively colored markers in Fig. 1(b). 
(b) Theoretical LDOS. Red lines represent states localized at the Fe atom (scaled by 1/10), while green lines correspond to states on neighboring Ta atoms. Gray-filled curve represents the total Ta states of the pristine surface for comparison. The inset shows the atomic structure of the on-center adsorbed DS cluster with the brown sphere representing the Fe atom.  
(c) Spin-resolved LDOS at the top layer. Gray-filled curve represents the total Ta states of the Fe adsorbed DS cluster. Red and blue filled curves represent spin-resolved contributions from the central Ta $5d_{z^2}$ orbital. Fe-related states (red and blue lines) are scaled by 1/6 for clarity.
(d) Charge density plots corresponding to the energy windows used in Fig.~1(j), highlighting the $d_{z^2}$-like character (top) and CDW modulations (bottom). In the bottom panel, the charge density is shown as the difference from the pristine surface.
(e) Experimental $dI/dV$ maps (top) and theoretical charge density plots (bottom) at energies corresponding to the Fe-related state and LHB, respectively. 
}
\end{figure}

% ---------- FIG 3
\begin{figure*}[t]
\centering{ \includegraphics[width=8 cm]{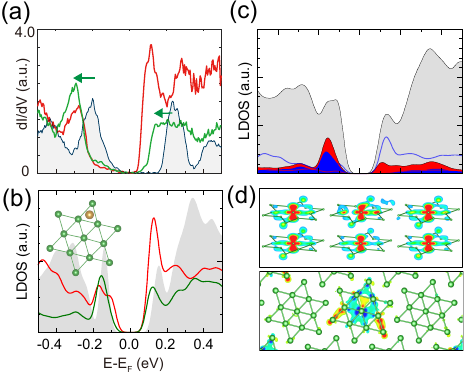} }
\caption{ \label{fig3}
On-edge Fe adsorption with weakly coupled charge-transfer.
(a) Point STS spectra taken at the Fe site (red), the CDW center beneath the Fe adatom (green), and a neighboring CDW center (filled with gray), corresponding to the respectively colored markers in Fig.~1(c). 
(b) Theoretical LDOS showing Fe-localized states (red) and the adjacent central Ta atom (green), with the pristine Ta LDOS shown for comparison (gray filled). The inset illustrates the on-edge DS adsorption geometry.  
(c) Spin-resolved LDOS at the top layer, showing the total Ta states (gray filled), spin-resolved contributions from the central Ta $5d_{z^2}$ orbital (red and blue filled curves), and Fe-derived states (red and blue lines, scaled by 1/3 for clarity). (d) Charge-density plots corresponding to the energy windows in Fig.~1(j), highlighting the $d_{z^2}$-like character (top) and CDW modulations (shown as charge difference, bottom).  
}
\end{figure*} 

% ---------- FIG 4

\begin{figure}[t]
\centering{ \includegraphics[width=8 cm]{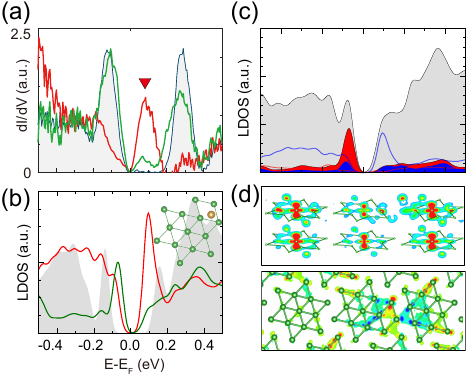} }
\caption{ \label{fig4}
Off-cluster Fe adsorption coupled to the CDW order.
(a) Point STS spectra taken at the Fe site (red), the CDW center beneath the Fe adatom (green), and a neighboring CDW center (gray filled), corresponding to the respectively colored markers in Fig.~1(d).  
(b) Theoretical LDOS showing Fe-localized states (red) and the adjacent central Ta atom (green), with the pristine Ta LDOS shown for comparison (gray filled). The inset illustrates the off-cluster DS adsorption geometry.  
(c) Spin-resolved LDOS at the top layer, showing the total Ta states (gray filled), spin-resolved contributions from the central Ta $5d_{z^2}$ orbital (red and blue filled curves), and Fe-derived states (red and blue, scaled by 1/3 for clarity). 
(d) Charge-density plots corresponding to the energy windows in Fig.~1(j), highlighting the $d_{z^2}$-like character (top) and CDW modulations (shown as charge difference, bottom).  
}
\end{figure} 

% ---------- FIG 5

\begin{figure}[t]
\centering{ \includegraphics[width=8 cm]{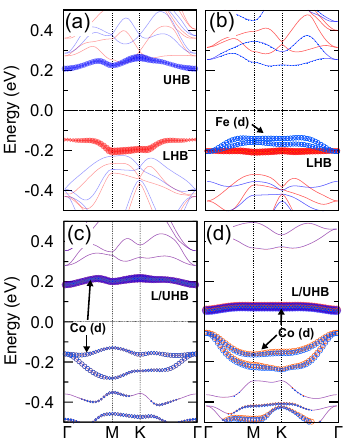} }
\caption{ \label{fig5}
DFT band structures for a on-center Fe or a on-center Co adatom on a single-layer 1T-TaS$_2$, using a $\sqrt{13}$$\times$$\sqrt{13}$ cell. (a) Pristine 1T-TaS$_2$ CDW structure for comparison. (b) Fe/1T-TaS$_2$. (c) Co/1T-TaS$_2$. (d) Co/1T-TaSe$_2$. Red (blue) filled circles indicate the majority (minority) spin states localized on Ta $d_{z^2}$ orbitals (LHB/UHB), while red (blue) open circles represent majority (minority) spin states localized on Fe and Co $d$ orbitals. The purple-colored bands in (c) and (d) represent spin-degenerate states, where the majority and minority spin states overlap.
}
\end{figure} 

\vspace{1em}

\vspace{1em}
%%%%%%%%%%%%%%%%%%%%%%%%%% DISCUSSION %%%%%%%%%%%%%%%%%%%%%%%%%%
\
\
\noindent{\fontfamily{phv}\selectfont \textbf{DISCUSSION}}

The major finding of the present work is that the direct chemical interaction between the local $d$ orbitals of Fe adatoms and the cluster Mott unit cell is dominating.
This effect, however, has been totally neglected in the previous discussion of exotic many-body effects.
For example, in a prior study on Co adsorption on 1T-TaSe$_2$~\cite{chen22}, the observed splitting of the LHB and UHB into two distinct peaks in STS respectively was attributed to a spinon Kondo effect, which hallmarks many-body coupling between localized magnetic moments and fractional spin excitations. 
However, in that work, neither the chemical nature of the Co impurity nor its orbital hybridization with the host surface was explicitly examined.
To revisit this interpretation, we extended our DFT analysis to Co adsorption on both 1T-TaS$_2$ and 1T-TaSe$_2$. Our calculations reveal that Co $3d$ ($3d^7$) orbitals hybridize significantly with both the LHB and the UHB, producing two-peak features remarkably similar to those reported experimentally (Fig. 5, Supplementary Fig. 11 and Fig. 12).
The dominating effect the direct orbital hybridization is shared in both Fe and Co cases, the detailed electronic structures is different (Supplementary Note II). The Co adsorption on both 1T-TaS$_2$ and 1T-TaSe$_2$ [Fig. 5(c) and 5(d)] shifts the LHB above the Fermi level and merges with UHB due to charge transfer to the Co atom. This process leaves even numbers of $d$ electrons on both Co and Ta atoms, eight and zero respectively, resulting in an overall spin-degenerate ground states. 
Consequently, there is no local spin and on LHB/UHB to induce spinon Kondo effects.
This suggests strongly that the previously observed spectral splitting in Co/1T-TaSe$_2$ also originates from chemical hybridization rather than from pure many-body effects.
In cases where a local magnetic moment survives, as in the Fe on-cluster configuration, many-body interactions may still be involved. However, because the present calculations successfully reproduce the experimentally observed spectral features, any many-body interactions, if present, are likely to play a secondary role, influencing only the overall size of the gap or level-splitting rather than generating distinct spectral signatures. 
Our study suggests that magnetic impurities can serve as a sensitive probe for different interactions within a cluster Mott system and that they may also act as a control knob for tuning such interactions separately.
Namely, the distinct site-dependent interactions identified here, including Mott gap suppression, charge-transfer-induced band renormalization, and the emergence of in-gap states, offer a way to manipulate the electronic structure at the atomic scale.

\vspace{1em}
%%%%%%%%%%%%%%%%%%%%%%%%%% Conclusion %%%%%%%%%%%%%%%%%%%%%%%%%%
\
\
\noindent{\fontfamily{phv}\selectfont \textbf{CONCLUSIONS}}

We have investigated the microscopic interactions of a magnetic impurity on the cluster-Mott system in the form of Fe adatoms the 1T-TaS$_2$ CDW phase using STM measurements and DFT calculations.
Our study reveals distinct adsorption sites of Fe atoms and their site-dependent interactions with the correlated CDW unit cells.
The on-center adsorption predominantly influences the Mott states through direct orbital hybridization, while the off-cluster adsorption perturbs the local CDW order emphasizing the two different aspects of a cluster Mott system, the localized correlated electrons and the structural degree of freedom within a unit cell. 
The on-edge adsorption, in contrast, exhibits a much weaker interaction, which simply dopes electrons into the CDW cluster.
For all three Fe adsorption configurations, the localized states induced by Fe observed in DFT calculations reasonably match the experimental STS features within the limitation in the accurate quantification of the correlation effect. 
The strong direct chemical interaction between the local $d$ orbitals of Fe adatoms and the cluster Mott unit cell rules out the possibility of subtle many-body interactions such as the spinon Kondo effect in explaining the observed in-gap states. 
These results indicate that CDW order and the Mott state may be independently tuned through the introduction of magnetic impurities, underscoring the promise of impurity engineering for manipulating these correlated phases in related materials.

\vspace{1em}
\noindent{\fontfamily{phv}\selectfont \textbf{METHODS}}

{\bf Scanning tunneling microscopy measurements.}
The STM measurements were conducted using a SPECS Joul-Thomson commercial cryogenic STM with Pt/Ir tips. All STM topography images were acquired at 4.5 K in constant-current mode. For scanning tunneling spectroscopy (STS), data were measured using a lock-in technique with a modulation amplitude of 10 mV and a frequency of 312 Hz.

{\bf Sample preparation.}
1T-TaS$_2$ single crystals were cleaved in high vacuum at room temperature. After cleaving, Fe atoms were deposited onto the 1T-TaS$_2$ sample using an e-beam evaporator (Focus GmbH, EFM3). Before deposition, the 1T-TaS$_2$ sample was cooled to 78 K. The deposition was then performed for 5 minutes with an e-beam voltage 1 kV and an emission current of 23.0 mA. After the deposition was complete, the 1T-TaS$_2$ sample was quickly mounted in the STM head and cooled down to 4.5 K.

{\bf Density-functional theory calculation.}
DFT calculations were performed using the Vienna \textit{Ab initio} Simulation Package (VASP)~\cite{kres96}, employing the Perdew--Burke--Ernzerhof (PBE) generalized gradient approximation (GGA) for the exchange-correlation functional~\cite{perd96}, together with the projector augmented-wave (PAW) method~\cite{bloc94}. A plane-wave energy cutoff of 400\,eV and an 8\,$\times$\,8\,$\times$\,1 Monkhorst-Pack \textit{k}-point mesh were used for the $\sqrt{13} \times \sqrt{13}$ unit cell. Atomic positions were relaxed until the residual forces were below 0.01\,eV/\AA. The $\sqrt{3} \times \sqrt{3}$ CDW structure was modeled using a TaS$_2$ bilayer slab with an in-plane lattice constant fixed at the equilibrium value of 3.374\,\AA, and a vacuum spacing of approximately 30\,\AA\ to avoid spurious interactions. The interlayer distance was fixed at the experimental value of 5.9\,\AA. To account for electron correlation effects, an on-site Coulomb interaction of $U = 2.3$\,eV was applied to the Ta 5$d$ orbitals, following Refs.~\cite{dara14, jung22}.

\
\

\noindent{\fontfamily{phv}\selectfont \textbf{ASSOCIATED CONTENT}}
%The Supporting Information is available free of charge at
%https://pubs.acs.org/doi/ /acsnano..

Further discussion on the band structures of pristine 1T-TaS$_2$ (Figure S1), the site-dependent properties of Fe adsorption (Figure S2-S10 and Note I), and the Co adsorption on the 1T-TaS$_2$ and 1T-TaSe$_2$ (Figure S11-S12 and Note II) (PDF)

\

\noindent{\fontfamily{phv}\selectfont \textbf{AUTHOR INFORMATION}}

\textbf{Jae Whan Park} - Center for Artificial Low Dimensional Electronic Systems, Institute for Basic Science (IBS), Pohang 37673, Republic of Korea

\textbf{Hyeonjung Kim} - Center for Artificial Low Dimensional Electronic Systems, Institute for Basic Science (IBS), Pohang 37673, Republic of Korea; Department of Physics, Pohang University of Science and Technology (POSTECH), Pohang 37673, Republic of Korea

\noindent{\fontfamily{phv}\selectfont \textbf{Corresponding Author}}

\textbf{Han Woong Yeom} - Center for Artificial Low Dimensional Electronic Systems, Institute for Basic Science (IBS), Pohang 37673, Republic of Korea; Department of Physics, Pohang University of Science and Technology (POSTECH), Pohang 37673, Republic of Korea; Email: yeom@postech.ac.kr	    % E-mail address for corresponding author

\
\

\noindent{\fontfamily{phv}\selectfont \textbf{Author contributions}}

H.W.Y. conceived the research idea and plan. J.W.P. performed the DFT calculations. H.K. carried out STM measurement. J.W.P., H.K. and H.W.Y. prepared the manuscript. J.W.P. and H.K. contributed equally to this work. 

\
\

%%%%%%%%%%%%%%%%%%%%%%%%%% Acknowledgement %%%%%%%%%%%%%%%%%%%%%%%%%%%%%
\noindent{\fontfamily{phv}\selectfont \textbf{ACKNOWLEDGEMENTS}}

This work was supported by Institute for Basic Science (Grant No. IBS-R014-D1).

\renewcommand{\refname}{\fontfamily{phv}\fontsize{12pt}{14pt}\selectfont

\end{document}